**The development of free-response questions to assess learning assistants' PCK in the context of questioning**

Beth Thacker[1], Stephanie Hart[2], Kyle Wipfli[1], Jianlan Wang[3]
[1]*Department of Physics and Astronomy, Texas Tech University, MS 41051, Lubbock, TX*
[2]*Terry B. Rogers College of Education and Social Sciences, West Texas A&M University, Canyon, TX*
[3]*College of Education, Texas Tech University, Lubbock, TX*

**Abstract:** We designed and developed a set of free-response questions to assess learning assistants' pedagogical content knowledge in the context of questioning (PCK-Q). Our goal was to develop a set of written free-response questions that would be representative of the LAs' PCK-Q observed in the classroom. In this paper we discuss the design, development and validation processes. These include videotaping and analysis of classroom interactions, development of a coding rubric, and development, administration, and validation of written free-response questions.

# I. INTRODUCTION

As introductory physics courses have increasingly focused on student engagement, there has also been an increase in reliance on the use of Learning Assistants (LAs) to lower the student-to-instructor ratio and increase active engagement with students in large classes. This is happening both in traditional and evidence-based course formats, such as SCALE-UP (Student-Centered Activities for Large Enrollment Undergraduate Programs) [1], ISLE (Investigative Science Learning Environment) [2], open-source tutorials, such Tutorials in Introductory Physics [3], INQ (Inquiry-based Experimental Physics) [4] and Modeling [5].

In this paper we focus on the evidence-based formats in the physics classroom. All of these formats are student-centered, use group learning environments, and are often experiment-based. As an instructional strategy, instructors use Socratic questioning [6] and guided-inquiry [7] to guide students through a more effective learning process than traditional lecture by asking questions [8, 9].

Significant research has demonstrated that instructors need to utilize components of pedagogical content knowledge (PCK) [10, 11] to implement these strategies effectively. Researchers have investigated some elements of PCK, including student understanding [12], elements of PCK for mathematical reasoning [13], impact of content knowledge for teaching [14-17].

In evidence-based formats, the instructors are not just faculty, but often include postdocs, graduate students and, as course sizes increase, undergraduate LAs. Therefore, LAs, as well as the other instructors, need to have sufficient PCK to effectively guide student learning. An important question that needs to be answered is: "What evidence do we have of LAs pedagogical content knowledge (PCK)?"

In order to answer this question, one needs to study the LAs' PCK. One way to do this is to videotape LAs in the classroom. The instructor overseeing the course can review the recordings with the LAs to discuss and assess their instructional practices. However, videotaping and reviewing the videotapes is a time-consuming and laborious process. There is a need for an efficient process to evaluate PCK other than direct observation through classroom visits or videotaping.

Since evidence-based strategies usually use Socratic questioning as a pedagogy, we decided to focus on developing a tool to evaluate LAs' PCK in the context of questioning pedagogies (PCK-Q). To accomplish this goal, we 1) videotaped LAs interacting with students in the classroom; 2) developed coding rubrics to evaluate LAs' PCK-Q based on the videotapes; 3) developed and validated a set of questions that can be used as an indicator of an LA's PCK-Q. Below, we present the development process, our coding rubrics, example questions, and validation process. We also discuss how the set of questions can be used to assess LAs' PCK-Q and help guide LAs towards more effective instruction.

# II. THEORETICAL FRAMEWORK

We used Magnusson's theoretical framework of pedagogical content knowledge (PCK) for science teaching. Magnusson describes PCK "as the transformation of several types of knowledge for teaching (including subject matter knowledge), and that as such it represents a unique domain of teacher knowledge" [10]. In their work, the five components of PCK are: (a) orientation toward science teaching, (b) knowledge of science curriculum, (c) knowledge of students' understanding of science, (d) knowledge of instructional strategies in science, and (e) knowledge of assessment in scientific literacy. Each component describes different sets of knowledge that are required for a science instructor's PCK. The components are based on Grossman [18] and Tamir [19] to expand upon their original definitions of PCK.

Magnussen's concept of "Orientation towards science teaching" (Orientation) refers to an instructor's knowledge of the purposes and goals for teaching at a particular level. Orientation refers to the overarching instructional practices which shape the different types of interactions and assessments in a course. Examples of orientations include students learning through guided inquiry, having students read different papers to discover the ideas of a certain concept, or lecturing. Orientation is a foundation for the other components of PCK and affects how different teachers' knowledge is shaped.

"Knowledge of science curriculum" (Curriculum Knowledge) has two major parts: (a) goals and objectives, and (b) specific programs and materials that are used in the curriculum. Goals and objectives of curricula not only include the content knowledge expected to be learned by students, but knowledge acquired from past learning experiences and the reasoning behind learning different aspects of content knowledge. Knowledge of specific programs and materials focuses on the awareness of different course designs and relevant materials, such as Physics Tutorials developed by the University of Washington [3] and Modeling Instruction [5]. The importance of knowledge of science curricula helps to distinguish instructors who are experts in both teaching and content from content experts.

"Knowledge of students' understanding of science" (Student Knowledge) refers to the instructor's awareness of students' content knowledge and how to improve students' understanding. Students' past learning experience impacts their present knowledge. Awareness of the students' knowledge and preconceptions impacts the way instructors choose to interact with their students. Awareness of common student difficulties prepares teachers to support students' learning. This part of PCK focuses on the awareness of students' conceptions rather than the curriculum knowledge itself.

"Knowledge of instructional strategies" (Instruction) refers to knowing both subject-specific and topic-specific strategies, used to teach science. Subject-specific strategies refer to the structure of a particular lab/dialogue/task to aid in students' acquisition of knowledge. Topic specific strategies refers to representations of content or principles to aid in students' acquisition of knowledge.

"Knowledge of assessment" (Assessment) in science has two interrelated parts: the methods used to assess students and the relative importance of assessing different dimensions of the curriculum. The assessment methods include both the awareness and implementation of established assessments, such as BEMA [20] and FCI [21], as well as assessments used within a course. Implementation refers to the format or the timing of the assessments. The knowledge of the importance of different dimensions refers to knowing which course objectives are important to evaluate. For example, knowledge of using different mathematical methods to solve for kinetic energy may be more important in calculus-based physics courses for STEM majors, while knowledge of how to use the conservation of energy in different simple contexts may be more important for algebra-based physics courses for non-STEM majors.

In the current research project, we interpreted Magnusson's definition of pedagogical content knowledge (PCK) for science teaching and the associated components to fit our context. We chose to focus our research on courses taught using guided inquiry rather than lecture-based instruction due to the prevalence in evidence-based course formats in physics. Our research specifically looks at PCK in the context of questioning, which narrows Magnusson's concept of Orientation (O) to looking at LAs' use of guided inquiry vs. direct instruction when discussing with students. We interpret Magnusson's knowledge of science curriculum (C) as LAs demonstration of implementation of classroom materials aligned with the purpose and design of the materials. In order for instructors to effectively use an evidence-based curriculum, instructors need to know how the curriculum was designed and its intended use. We define Student Knowledge (S) as an LA's ability to identify students' strengths and difficulties while engaging in discussion. We narrow Magnusson's concept of Instruction (I) to LAs' use of guiding techniques that have been demonstrated to be effective when interacting with students. Because LAs do not normally create or grade assessments, we did not include Magnussen's component of Assessment (A) in this project. We refer to our adaptation of Magnusson's work as the OCSI (Orientation/Curriculum/Students/Instruction) framework.

### III. SETTINGS AND PARTICIPANTS

We collected data at two universities, an algebra-based course at Texas Tech University (TTU) and a calculus-based course at Florida International University (FIU). Data was collected at TTU in Fall 2019 and Spring 2020. Due to the Covid pandemic and the cancellation of in-person classes, data collection at TTU was stopped in March 2020. In Fall 2020 and Spring 2021, we collected data at FIU. These courses were being taught online via Zoom.

| Component | Magnusson's Definition | Magnusson's definition applied in our context |
|---|---|---|
| Orientation | Knowledge of the overarching instructional practices which shape the different types of interactions and assessments in a course. | LAs' use of guided inquiry vs. direct instruction when discussing with students. |
| Curriculum | Knowledge of goals and objectives, and specific programs and materials that are used in the curriculum. | LAs' demonstration of implementation of classroom materials aligned with the purpose and design of the materials. |
| Students | Awareness of students' content knowledge and how to improve students' understanding. | LAs' ability to identify students' strengths and difficulties when engaging in discussion. |
| Instruction | Knowledge of subject-specific and topic-specific strategies, used to teach science. | LAs' use of guiding techniques that have been demonstrated to be effective when interacting with students. |
| Assessment | Knowledge of the methods used to assess students and the relative importance of assessing different dimensions of the curriculum | Not applicable |

FIG. 1. Comparison of Magnusson's components of PCK and our use of PCK when analyzing LAs.

### A. Algebra-based course

The algebra-based introductory physics sequence at TTU consists of two semesters, the first primarily covers mechanics and the second primarily covers electricity and magnetism (E&M). The student population is predominantly health science majors, including premedical, predental, pre-physical therapy, etc. The number of students registered for the first semester is usually about 400 and around 250 students register for the second semester. Each semester, there are two INQ sections of 60 students each. The rest of the students are in lecture-based sections of 60-200 students depending on instructor schedules and classroom availability. This study focused on the INQ sections only.

The INQ section was developed with National Science Foundation (NSF) funding 20 years ago and has been taught as one or more section(s) of the algebra-based course every semester since then. It was developed explicitly for health science majors, taking their needs, learning styles, backgrounds and motivations into account. It is taught without a required textbook in a studio-style environment with Socratic questioning pedagogy. Students work through the units in groups, learning to develop both quantitative and qualitative models based on their observations and inferences and then using the models to make predictions and solve problems. The goal is to start in the laboratory like scientists and learn the content through guided experimentation. The materials consist of the laboratory units, pretests, readings, and exercises. There are also homework sets, exams, and quizzes. The course covers approximately the same content as is covered in the other sections of the class, but with more of a focus on developing models based on experimentation and on developing observational, analytic, and thinking skills. Students can choose to be in either the INQ or traditional section, but most often it is determined by what fits their schedules and when they need to take the course, as far as graduation. The LAs are chosen from students who have taken the course before in the INQ format. Each course had 2-3 LAs, 1-2 graduate students (TAs), and an instructor.

### B. Calculus-based course

Similar to TTU, the FIU sequence of two physics courses consists of the first covering mechanics and the second covering E&M. The sequence taught at FIU is a calculus-based Modeling [5] sequence, where students work in groups to construct physics knowledge. In the two semesters we collected data, the instructor and LAs met with students online via Zoom, where students worked in break-out rooms. Hands-on physics experiments were replaced with virtual labs. The LAs moved around different breakout rooms to facilitate student exploration. Towards the end of a lab, all the students returned to the main room on Zoom to reflect on their exploration and summarize the results.

### C. LAs

*i. TTU*

At TTU, the participants were four LAs and one graduate student teaching assistant (TA) in the introductory physics sequence. LA-1 was a Hispanic male undergraduate student, who had taken the INQ sequence and had no prior teaching experience. LA-2 was a white female undergraduate student, who also took the INQ sequence with no prior teaching experience. LA-3 was a Hispanic female undergraduate physics major, who had a traditional lecture-based sequence and not the INQ sequence. LA-4 was a white male undergraduate student, who had taken the INQ sequence with no prior teaching experience. The TA was a Middle Eastern female graduate student, who had 1 year of experience teaching the INQ course. The instructional team for the INQ course had weekly preparatory sessions that covered both content and pedagogy.

*ii. FIU*

At FIU, the participants were three LAs in the introductory physics sequence. LA-1 was a Hispanic female undergraduate student who had no experience serving as an LA prior to this study. LA-2 was a black female undergraduate student who had 1.5 years of experience serving as an LA in introductory physics courses. LA-3 was a Hispanic male undergraduate student who had 1.5 years of experience serving as an LA in introductory physics courses. Before this project, the three LAs had taken an instructional methods course regarding pedagogy for inquiry teaching such as scaffolding and questioning. During this course, the LAs met with the instructor online during weekly preparation sessions to preview physics content knowledge for the coming week.

## IV. METHODS

In this section we describe the multiple stages of data collection and data analysis. To do this we followed a modified version of the framework by Adams & Weiman [22]. First, we recorded student-LA interactions in the introductory physics classrooms. Second, we analyzed video analysis and developed rubrics to describe interactions between LAs and students. Third, we verified interrater reliability. Fourth, we developed the free-response questions based on authentic interactions from the videos. Last, we validated the free-response questions by using the same rubric to code both the free-response and the videos.

### A. Classroom Video Recording

At TTU, we video recorded the INQ sequence for two semesters. We used a video capture system (Swivl) [23] for video recording of student-LA interactions during class periods. The LAs and instructors wore microphones and the Swivl device pivoted to follow the LAs around the room and record video of student-LA interactions.

At FIU, we video recorded the Modeling sequence on Zoom. LAs recorded their breakout room sessions and submitted the recordings to us.

By recording we were able to analyze the various types of interactions that would come about during a normal class period and to categorize the different abilities of LAs based on the types of questions they asked.

### B. Video Analysis and Rubric Development

We analyzed multiple videos per week of each LAs' interactions. We examined the video recordings sentence by sentence and identified the different types of statements that LAs would make when interacting with the students, grouping them into categories. We called this our microscopic coding system. As we applied the microscopic coding system to multiple video recordings, we recognized patterns in the LAs statements and categorized them into levels that corresponded with the components of PCK. We called this our macroscopic coding system.

*i. Microscopic coding system*

We observed that LAs generally used a tutoring-type interaction or a questioning-type interaction. Tutoring-type interactions involve an LA explaining concepts to students or walking them through problem solving step-by-step by providing information. Questioning-type interactions involve an LA primarily using questions rather than statements to guide students to a correct conceptual understanding of concepts or through problem solving. We identified three types of questioning interactions, which we labeled guiding questions, probing questions and checking questions.

Guiding questions are questions that guide students' thinking by helping them to reason through to a conclusion while evaluating experimental evidence, solving a problem or coming to a correct conceptual understanding. We were particularly interested in a sequence of guiding questions because it usually demonstrates all four elements of PCK-Q: content knowledge, knowledge of the curriculum, knowledge of students' understanding and knowledge of instructional strategies to address student difficulties.

Probing questions are questions used to collect information about students' understanding.

We observed that interactions often began with probing questions. They are questions like "Why did you drop it from that height?" or "Could you tell me why you placed the magnet here?" They are to collect further information about the experimental set-up or possible student difficulties, for example.

Checking questions (also called temperature checking) are simple questions, usually to be answered by yes or no, about the experimental set-up or student understanding. Examples are "Did you zero the sensor?" or "Does this make sense to you?"

The microscopic coding scheme is shown in Fig. 2.

| Category | Code | Description | Examples |
| --- | --- | --- | --- |
| Guiding Question | gq | LAs ask questions that guide students thinking to help them to reason through to a conclusion while evaluating experimental evidence, solving a problem or coming to a correct conceptual understanding. | What happens if you pull with a constant force, say 1 unit of rubber band force? If you pull on a cart on a frictionless track with one unit of rubber band force, what happens? Could you make the same motion with a bungee cord? How? How would you compare the two? |
| Probing Question | pq | LAs ask questions to collect information about students' understanding. | 1) Why did you drop it from that height? 2) Could you tell me why you placed the magnet here? |
| Checking Question | cq | LAs ask general questions to collect information (temperature checking). | 1) Did you zero the sensor? 2) Does this make sense to you? |
| Tutoring/ Lecturing | le | LAs imparts information directly to the students. | The gravitational force would not be the same because the masses are different. |

FIG. 2. Microscopic coding scheme with descriptions and examples of each code. The table above includes the code category, the label for the code, the description of the code itself and an example statement or question from each category.

### ii. Macroscopic coding

In our initial (microscopic) coding system, we coded individual questions/statements made by LAs. Over time, we observed that LAs often used a series of questions and/or statements in order to guide students to a correct conceptual understanding or help them solve a problem. We established the term "vignette" to describe a period of time from when an interaction began to when an LA ended the interaction with that group of students.

Within each vignette, we identified four main patterns in the microscopic coding that could be organized into levels. We observed some vignettes where LAs used a repeated series of probing and guiding questions. The LAs guided students through a thought process or an experiment using a sequence of guiding questions interspersed with probing and checking questions. We also observed vignettes where LAs used mostly probing and checking questions interspersed with sections of lecturing. It appeared that the intent of the LA was to guide students through questioning, but at some point(s) the LA shifted to tutoring or lecturing. There are usually one or two guiding questions, but also a significant amount of lecturing/tutoring. Because these two groups employed questioning, we categorized these groups as a Questioning orientation. The first group demonstrated the most skill at questioning, so we labeled the pattern Qa. The second group used questioning less frequently, so we labeled this pattern Qb.

| Level | Description | Coding Pattern |
|-------|-------------|----------------|
| Qa | LAs guide students through a thought process or an experiment using a sequence of guiding questions interspersed with probing and checking questions. This is the most direct evidence of PCK-Q. | pq-gq-(pq/tr/cq)-gq-(pq/tr/cq) very little le |
| Qb | It appears that the intent of the LA is to guide students through questioning, but at some point(s) the LA shifts to tutoring or lecturing. There are usually one or two guiding questions, but also a significant amount of lecturing/tutoring. This provides some evidence of PCK-Q, but not as much as Qa. | (pq/cq)-gq-le- is the general form. This may repeat. |
| Qe | Qa or Qb with an identifiable content error. | er |
| Da | Tutoring -- lecturing with some probing and checking questions. These LAs may have PCK, but not PCK-Q. | series of pq-le-cq |
| Db | Basically lecturing. The LA does not take students' understanding into account. Not much evidence of PCK or PCK-Q, except for C, content knowledge. | cq-le |
| De | Da or Db with an identifiable content error. | er |

FIG. 3. Macroscopic coding scheme with descriptions and coding patterns. The table above includes the label for each level, the description of the level itself and an example of a pattern that indicates that the interaction should be coded by that level.

In contrast to the questioning groups, we also observed vignettes where LAs primarily lectured or tutored to guide students, with some probing and checking questions. We labeled this pattern Da. At times, we observed vignettes where LAs lectured or tutored only asking one or two checking questions. It appeared that the intent of the LA was to impart information without taking the students' understanding into account. We labeled this pattern Db.

If a vignette would have been labeled as Qa or Qb but contained an identifiable physics error, it was coded Qe. Similarly, if a vignette would have been labeled as Da or Db but contained an identifiable physics error, it was coded as De.

Each videotaped recording contained multiple vignettes. Each vignette was coded as Qa, Qb, Da, Db, Qe, or De. Within a single videotaped recording, an LA demonstrated one or more levels. The macroscopic coding scheme is shown in Fig 3.

### iii. Revisiting the OCSI (Orientation/Curriculum/Students/Instruction) framework

We recognized that the levels of Qa, Qb, Da, and Db are related to Magnussen's components of PCK applied in our context. To clarify the relationship, we also coded the videotape recordings using the OCSI framework. We matched the OCSI framework to the levels of the macroscopic coding scheme, as shown in Fig. 4.

The level labeled Qa represents a situation in which the LA guides students through a thought process or an experiment using a sequence of guiding questions interspersed with probing and checking questions. This is the most direct evidence of all four components of PCK: Orientation, Knowledge of Curriculum, Knowledge of Students and Knowledge of Instructional Strategies (OCSI). It involves at least one, but usually a series of guiding questions.

The level labeled Qb represents a situation in which it appears that the intent of the LA is to guide students through questioning, but at some point(s) the LA shifts to tutoring or lecturing. There are usually one or two guiding questions, but also a significant amount of lecturing/tutoring. This provides evidence of three components of PCK: Orientation, Knowledge of Content, and Knowledge of Students (OCS). Qb demonstrates a questioning orientation but does not show evidence of effective instructional strategies, such as a sequence of guiding questions (I).

The level labeled as Da represents a situation where an LA is primarily tutoring or lecturing. The LA may use occasional probing and/or checking questions, but the conversation is dominated by statements. The components of PCK evidenced by Da are Knowledge of Content and Knowledge of students (CS). Unlike the Qa, Qb and Qe levels, the Da, Db and De levels do not show evidence of a questioning orientation (O and I).

The level labeled as Db represents a situation where an LA is tutoring or lecturing without taking the students' understanding into account. The only component of PCK that is evidenced by Db is Knowledge of Content (C).

|    | O | C | S | I |                             |
|----|---|---|---|---|-----------------------------|
| Qa | X | X | X | X | x = more likely to correlate |
| Qb | X | X | X | - | - = less likely to correlate |
| Da | - | X | X | - |                             |
| Db | - | X | - | - |                             |

FIG. 4. Correlation between macroscopic coding levels and components of OCSI (Orientation/ Content /Students /Instruction).

### C. Interrater reliability for video analysis

We determined inter-rater reliability across our coding of the Qa, Qb, Da, and Db levels by choosing data from one week of Spring 2020 recordings from TTU. To determine inter-rater reliability, three raters independently coded the vignettes of two LAs using the macroscopic coding system. The unweighted Cohen's kappa for comparison of any two raters was always above 0.75 and the weighted Cohen's kappa was always above 0.86. These values indicate substantial to near perfect agreement.

### D. Free Response Questions Development

Questions were developed based on classroom interactions. Each question stem consisted of the contextual background, a description and/or dialogue of a student-LA interaction, and open-ended questions for a respondent to answer. To write the stems we chose student-LA interactions from the video recordings or classroom observations that exemplified common student difficulties. We described the interactions in the stem of the questions providing enough detail to illustrate the students' difficulty, sometimes using direct quotes. The questions are intended to reflect authentic student interactions in the classroom and therefore contain student work, student dialogue, errors, and misconceptions. We show examples of the questions in Figs. 5 and 6. Additional sample questions can be found in Appendix A.

We used an iterative process to refine the open-ended questions that produced LA responses that were consistent with their video interactions in the classroom. Initially, the answers from LAs were too vague and did not reflect the level of PCK observed in the videos. Therefore we refined the open ended questions by explicitly asking about specific components of PCK to better elicit components of PCK as demonstrated in the videos.

The final result was a two-part prompt, as illustrated in the example in Figs. 4 and 5. Part (a) asks LAs to infer students' strengths and difficulties from the scenario in the question stem. It was specifically designed to measure the Knowledge of Students, the S component of OCSI. Part (b) asked respondents to describe how they would respond to the students and why. It assesses the remaining components of PCK-Q: Orientation, O, Content Knowledge, C, and Knowledge of Instructional Strategies, I.

---

**Context: Students have studied Newton's Laws. They are now studying collisions and trying to develop an equation relating impulse and change in momentum.** The students are taking two carts on a frictionless track and colliding them together. Each run of the carts, the carts have different masses. Probe sensors are attached to find how much force acts on each cart in the collision.
The students' results are in the table below.

| Mass of Cart 1 | Mass of Cart 2 | Force on Cart 1 | Force on Cart 2 |
|---|---|---|---|
| 1.00 kg | 1.00 kg | 7.41 N | 7.41 N |
| 1.25 kg | 1.00 kg | 11.38N | 12.38 N |
| 1.50 kg | 1.00 kg | 15.10 N | 16.10 N |

The students conclude that the more mass the cart has the more force the cart exerts in the interaction.

a. What can you conclude from the information provided about the students' content knowledge? What are the students' strengths and difficulties?
b. How would you respond to the students? Please use direct quotes of what you would say. What is the reasoning behind your response?

FIG. 5. Example Classical Mechanics question.

**Context: Students are studying magnetism. They are learning about magnetic fields and forces.**

Students are shown two wires carrying current in opposite directions and asked to indicate the direction of the magnetic field at wire 2 due to wire 1 in the plane of the paper and then use that to determine the direction of the magnetic force on wire 2. They draw the following:

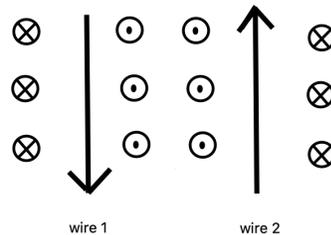

wire 1     wire 2

They are confused because they don't know from their picture the direction of the field of wire 1 at wire 2 and therefore cannot determine if the force is to the left or to the right. What is the students' difficulty and how should LA address it?

   a. What can you conclude from the information provided about the students' content knowledge? What are the students' strengths and difficulties?
   b. How would you respond to the students? Please use direct quotes of what you would say. What is the reasoning behind your response?

FIG. 6. Example Electricity and Magnetism question.

### E. Free Response Questions Analysis

The free response questions were administered in sets of 3-5 questions 4-5 times throughout the semester at TTU. At FIU, 1-2 questions were administered each week. The free response questions were administered roughly during the time period the concepts were addressed in class. We coded the responses based on the macroscopic coding rubric patterns. Each of the LA's responses were then assigned a level of the macroscopic coding rubric.

### F. Inter-rater reliability for free-response question analysis

We established inter-rater reliability across our coding of the Qa, Qb, Da, and Db levels for the free-response questions. For this we used the data collected in Spring 2020 from TTU. To determine inter-rater reliability, three raters independently coded 15 of the LAs' written free-response answers using the macroscopic coding system. We then calculated Cohen's kappa for pairs of raters. The unweighted Cohen's kappa for comparison of any two raters was always above 0.69 and the weighted Cohen's kappa was always above 0.78. These values indicate substantial to near perfect agreement.

### IV. VALIDATION

We demonstrated criterion validity of the free-response questions by comparing the video coding to the free-response coding for each LA. For the videos, we randomly selected one day of vignettes each week and documented the percentage of occurrences for each PCK level for that day. Then we averaged the percentage of occurrences of each level over all the weeks. For all the free-response questions administered we determined the percentage of occurrences for each level. We then compared the video and written free-response results. This is shown in Fig. 7 for TTU and in Figs. 8 and 9 for FIU.

We found reasonable agreement between the video and written free-response responses. For example, for student AF, the written responses were mostly Qa and Qb and the video was also mostly Qa and Qb. For student TZ, both the video and the written free-response questions were mostly Qb and Da. In Fall 2020, the LAs only worked 2-3 written free-response problems so the trends were not as significant because of minimal written free-response data.

| TTU | | Qa | Qb | Da | Db | Qe/De |
|---|---|---|---|---|---|---|
| AF | Video | 29% | 51% | 17% | 3% | 0% |
| | Written | 50% | 25% | 0% | 0% | 25% |
| TZ | Video | 4% | 29% | 60% | 4% | 2% |
| | Written | 0% | 60% | 20% | 20% | 0% |
| HB | Video | 6% | 10% | 84% | 0% | 0% |
| | Written | 0% | 20% | 40% | 40% | 0% |

FIG. 7. Average occurrence of PCK-Q levels over multiple vignettes for TTU students.

| FIU Spring | | Qa | Qb | Da | Db | Qe/De |
|---|---|---|---|---|---|---|
| VF | Video | 41% | 3% | 50% | 3% | 3% |
| | Written | 63% | 0% | 25% | 0% | 12% |
| DD | Video | 11% | 19% | 51% | 11% | 8% |
| | Written | 0% | 0% | 73% | 0% | 27% |

FIG. 8. Average occurrence of PCK-Q levels over multiple vignettes for FIU students during Spring 2021.

| FIU Fall | | Qa | Qb | Da | Db | Qe/De |
|---|---|---|---|---|---|---|
| SL | Video | 3% | 10% | 57% | 14% | 14% |
| | Written | 33% | 67% | 0% | 0% | 0% |
| DD | Video | 0% | 11% | 78% | 11% | 0% |
| | Written | 0% | 0% | 50% | 0% | 50% |
| AM | Video | 17% | 33% | 17% | 33% | 0% |
| | Written | 33% | 67% | 0% | 0% | 0% |

FIG. 9. Percentage of occurrences for PCK-Q levels over multiple free-response questions for FIU students during Fall 2020.

## V. CONCLUSIONS AND FUTURE

In this project, we created a set of questions that can be used to measure PCK-Q and demonstrated criterion validation showing that the questions are able to predict LAs' level of PCK-Q observed in classroom videos. This addresses the need for an efficient means for measuring PCK-Q. These questions can be used to inform instructors about their LAs' PCK-Q levels in the classroom.

In this study, we encountered some limitations due to small sample size and the number of questions answered. The current study was limited by a small sample size of LAs in introductory, laboratory-based physics classes. Another limitation we observed was that our ability to predict LA's PCK-Q was limited for LAs who answered less than three questions.

The questions we developed can be used for assessment or for LA training purposes. Questions can be administered in a single set or 1-2 per week. Instructors could use the questions to assess their LAs. Alternatively, the instructor could use the questions for training LAs during class preparation sessions.

In addition, for widespread distribution, it may be beneficial to have a set of easy-to-score questions using formats such as multiple choice or Likert-style. These formats would take less time to administer, be faster to score, and would not require trained coders. Presently, we are in the process of developing Likert-style questions.

## VI. ACKNOWLEDGEMENTS

This work was supported by the National Science Foundation Improving Undergraduate Science Education Grant 1838339.

---